\documentclass{iaus}
\usepackage{graphicx}

\title[Helium-rich stars in $\omega$\,Cen] 
{On the origin of the helium-rich population in the peculiar globular cluster 
Omega Centauri}

\author[D. Romano et al.]   
{D. Romano$^{1, 2}$, 
M. Tosi$^2$, M. Cignoni$^{1, 2}$, F. Matteucci$^3$, E. Pancino$^2$ 
\and M. Bellazzini$^2$}

\affiliation{$^1$Dept. of Astronomy, Bologna University, \\ Via Ranzani 1, 
I-40127, Bologna, Italy \\ email: {\tt donatella.romano@oabo.inaf.it} 
\\[\affilskip]$^2$INAF-Bologna Observatory, \\ Via Ranzani 1, I-40127, Bologna, 
Italy
\\[\affilskip]$^3$Dept. of Physics, Trieste University, \\ Via Tiepolo 11, 
I-34143, Trieste, Italy}

\pubyear{2010}
\volume{268}  
\jname{Light elements in the Universe}
\editors{C. Charbonnel, M. Tosi, F. Primas \& C. Chiappini, eds.}
\begin{document}

\maketitle

\begin{abstract}
In this contribution we discuss the origin of the extreme helium-rich stars 
which inhabit the blue main sequence (bMS) of the Galactic globular cluster 
Omega Centauri. In a scenario where the cluster is the surviving remnant of a 
dwarf galaxy ingested by the Milky Way many Gyr ago, the peculiar chemical 
composition of the bMS stars can be naturally explained by considering the 
effects of strong differential galactic winds, which develop owing to multiple 
supernova explosions in a shallow potential well.
\end{abstract}

\firstsection 
\section{A dwarf galaxy progenitor suffering strong galactic winds}

\begin{figure}[h!]
  \begin{center}
  \includegraphics[height=10cm]{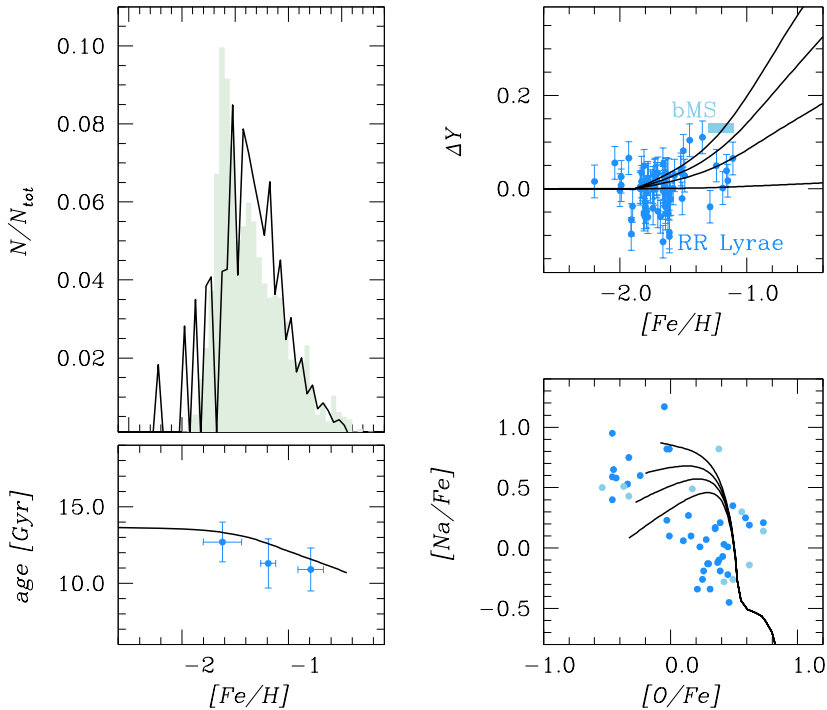} 
  \caption{Predicted (thick solid lines) MDF (top left-hand panel), AMR (bottom 
    left-hand panel), relative He enrichment (top right-hand panel) and Na-O 
    anticorrelation (bottom right-hand panel) for $\omega$\,Cen stars compared 
    to observations from \cite{s05} shaded histogram, top left-hand panel), 
    \cite{h04} dots with error bars, bottom left-hand panel), \cite{s06} dots, 
    top right-hand panel), \cite{n04} box, top right-hand panel), \cite{ndc95} 
    and \cite{s00} (dots, bottom right-hand panel). The upper (lower) curves in 
    the right-hand panels correspond to models computed with lower (higher) 
    efficiencies of He and Na entrainment in the galactic outflow (see 
    \cite{r10}).}
  \label{fig1}
 \end{center}
\end{figure}

It has been suggested -- and it is now widely accepted -- that the kinematical, 
dynamical and chemical properties of the most massive globular cluster of the 
Milky Way, Omega Centauri ($\omega$\,Cen, NGC\,5139), can all be understood if 
it is the surviving remnant of a larger system, captured and partially 
disrupted by the Milky Way many Gyrs ago (see, e.g., \cite{r07}, and references 
therein). However, the origin of the extreme He-rich stars hosted on its blue 
main sequence (bMS) still awaits a satisfactory explanation. Here we propose a 
possible solution, in the framework of a chemical evolution model which 
reproduces other major observed properties of the cluster, namely, its stellar 
metallicity distribution function (MDF), age-metallicity relation (AMR), trends 
of several abundance ratios with metallicity and Na-O anticorrelation 
(Fig.~\ref{fig1}; see \cite{r07}, 2010).

In order to reproduce all the relevant observations, the parent galaxy must 
experience both infall of gas of primordial chemical composition and outflow 
of processed matter. The key assumption that allows the formation of extreme 
He-rich stars is that, while supernova (SN) ejecta are efficiently lost through 
the galactic outflow, elements restored to the interstellar medium (ISM) by 
gentle winds from both asymptotic giant branch (AGB) and fast rotating massive 
stars (FRMSs) are mostly retained in the cluster potential well (see \cite{r10} 
and references therein for details).

He and Na are dispersed in the ISM through slow stellar winds by both AGBs and 
FRMSs. O, instead, as well as other heavy elements, is expelled by massive 
stars through fast polar winds. Hence, in the framework of our model, He and Na 
are preferentially retained inside the cluster potential well, while O is 
mainly vented out through the galactic outflow. This naturally produces the 
Na-O anticorrelation and He-rich bMS population, as observed (Fig.~\ref{fig1}, 
right-hand panels). The coexistence of populations with `normal' and `high' He 
content at [Fe/H]~$\ge -$1.4 is expected if He is removed with different 
intensities from different regions of the proto-cluster.

\firstsection

\end{document}